\documentclass[prb,onecolumn,showpacs]{revtex4}
\usepackage{color,soul}
\usepackage{amsmath}
\usepackage{dcolumn}
\usepackage{graphicx}
\usepackage{bm}
\usepackage{amssymb}
\usepackage{subfigure}
\begin{document}

\title{Why Matthiessen's rule is violated in the high-$T_{c}$ cuprate superconductors?}
\author{Xinyue Liu and Tao Li}
\affiliation{Department of Physics, Renmin University of China, Beijing 100872, P.R.China}

\begin{abstract}
The perfect linear-in-$T$ dc resistivity in the strange metal phase of the high-$T_{c}$ cuprate superconductors is probably the most prominent manifestation of their non-Fermi liquid nature. A major puzzle about the strange metal behavior is that there is no discernible change in the trend of the dc resistivity at a temperature when we expect the electron-phonon coupling to play an essential role. The empirical Matthiessen's rule of summation of the resistivity from different scattering channels seems to be simply violated. On the other hand, the electron-phonon coupling has long been proposed to be responsible for the various spectral anomalies observed in the cuprate superconductors. In particular, the $B_{1g}$ buckling mode of the oxygen ion in the $CuO_{2}$ plane is proposed to be responsible for the peak-dip-hump spectral shape around the anti-nodal point. Here we show that the coupling to the $B_{1g}$ buckling mode is strongly suppressed by the vertex correction from the antiferromagnetic spin fluctuation in the system as a result of the destructive interference between electron-phonon matrix element at momentum differ by the antiferromagnetic wave vector. In the real space scenario, such a destructive interference effect simply amounts to the fact that the electron hopping between nearest-neighboring sites is suppressed by the antiferromagnetic spin correlation between the electrons. More generally, we argue that the electron-phonon coupling strength in the cuprate superconductors should diminish with the proximity of the Mott insulating phase as a result of similar vertex correction from the electron correlation effect. We think that this offers an interpretation for the violation of the Matthiessen's rule in the dc resistivity of the strange metal phase of the cuprate superconductors.       
\end{abstract}

\maketitle

The dc resistivity of the high-$T_{c}$ cuprate superconductors is found to exhibit perfect linear-in-$T$ behavior from very high temperature all the way down to zero temperature around the pseudogap endpoint\cite{Cooper,Taillefer,Proust,Hussey,Phillips}. Such an anomalous behavior is hard to reconcile with the fermi liquid transport picture in which the dc resistivity is determined by the quasiparticle scattering rate on the fermi energy. For example, in a conventional metals with a phonon-dominated resistivity, a $T^{5}$ dependence in $\rho(T)$ is expected in the low temperature regime as a result of the suppression of the thermal excitation of the phonon mode contributing to the momentum relaxation of the electron. At higher temperature, $\rho(T)$ crossovers to a linear behavior at a temperature of the order of the Debye frequency $\omega_{D}$. Such an evolution is beautifully illustrated in the $T/\omega_{D}$ scaling of the dc resistivity.  

It is widely speculated that some kind of quantum critical fluctuation is at the origin of such a strange metal behavior. However, it is still not generally accepted if there is indeed such a genuine quantum critical point(QCP) in the phase diagram of the cuprate superconductors that is responsible for the strange metal behavior. Even if we admit such a QCP scenario, it is still puzzling why the slope of the dc resistivity remains its low temperature value at a temperature when we expect the electron-phonon scattering to play an essential role. In fermi liquid metals, the quasiparticle scattering rate contributed by different scattering channels usually add up to determine the dc resistivity. More specifically, we usually expect that 
\begin{equation}
\frac{1}{\tau}=\frac{1}{\tau_{imp}}+\frac{1}{\tau_{e-e}}+\frac{1}{\tau_{e-p}}
\end{equation}
in which $1/\tau$ is the total scattering rate of the electron, $1/\tau_{imp}$, $1/\tau_{e-e}$ and $1/\tau_{e-p}$ denote respectively the scattering rate contributed by impurity, electron-electron and electron-phonon scattering. According to the Drude model, this leads to the addition of the dc resistivity from different scattering channels. This is the so-called Matthiessen's rule. Clearly, such a rule is violated in the strange metal phase of the cuprate superconductors. While such a violation may simply imply the failure of the quasiparticle transport picture\cite{Hartnoll}, here we want to know if such a violation is possible within such a traditional transport picture. 

The phonon modes and their coupling to the electron have been extensively studied in the cuprate superconductors\cite{RLiu,Ginsberg,TPD,Gunnarsson,Nagaosa,Shen,Peng}. The phonon modes most relevant for the electron moving in the $CuO_{2}$ plane are those related to the vibration of the in-plane oxygen ions, which can be classified into the breathing mode moving along the $Cu-O$ bond direction and the buckling mode moving perpendicular to it. It has been argued that these in-plane oxygen modes are responsible for the various spectral anomalies observed in ARPES measurements. More specifically, the coupling to the breathing mode at $70\ meV$ is argued to be responsible for the dispersion kink in the nodal direction. Similarly, the coupling to the $40\ meV$ oxygen buckling mode is believed to be responsible for the peak-dip-hump structure in the anti-nodal spectrum\cite{Damascelli,He,Zhou}. More recently, it is claimed that the coupling strength between the $B_{1g}$ buckling mode and the anti-nodal quasiparticle suffers from a sudden drop around the pseudogap endpoint upon doping\cite{He}. In the single band picture, the coupling of these modes to the electron can be roughly classified into two categories, namely, the diagonal coupling to the on-site electron density and the off-diagonal coupling to the inter-site electron hopping. Since charge density fluctuation is suppressed at low energy in a doped Mott insulator, it is believed that the off-diagonal electron-phonon coupling is playing a more important role\cite{Nagaosa}. A particular example of the electron-phonon coupling in the off-diagonal channel is the coupling to the $B_{1g}$ buckling mode. Indeed, both theoretical and experimental investigations indicate that the coupling to the $B_{1g}$ buckling mode at about $40 \ meV$ is the most relevant for the cuprate physics\cite{Shen,He}. At the same time, strong electron-phonon coupling has been invoked to understand the extremely broad spectral shape of a single hole moving in the Mott insulting  background\cite{Damascelli,KMShen,Nagaosa1} for parent compound of the cuprate superconductors.

However, the actual role of the electron-phonon coupling in the cuprate superconductors is still under debate. On the theoretical side, first principle calculation indicates that the electron-phonon coupling strength in the cuprate superconductors is order of magnitude smaller than needed to interpret the observed spectral anomalies in ARPES measurements\cite{Louie,Heid}. It is argued that such a discrepancy may have its origin in the strong correlation effect which is neglected in the first principle calculation\cite{Reznik}. However, exactly how the electron correlation effect would affect the electron-phonon coupling in the cuprate superconductors is not clear. Naively, the suppression of the electron itinerancy by the strong correlation effect may suppress the Coulomb screening and thus enhance the bare electron-phonon coupling strength. On the other hand, the strong electron incoherence induced by the correlation effect is expect to suppress the bare electron-phonon vertex. The suppression of the charge fluctuation near the Mott insulating phase may also render the coupling to the electron density irrelevant at low energy\cite{Nagaosa}. Indeed, recent ARPES measurement on multi-layer cuprate finds that the quasiparticle excitation in a Mott insulating background can be as sharp as that observed in optimally doped cuprates, provided that the doping induced disorder effect is properly screened\cite{Kondo}. This indicates that the polaron effect in the parent compound of the cuprate superconductors is orders of magnitude weaker than that estimated before. Finally, we note that certain phonon branch may suffer strong renormalization when the electron subsystem is in proximity to a charge density wave(CDW) ordering instability. While this is indeed observed in the RIXS measurements of some cuprate superconductors\cite{Comin}, we think that this should be better interpreted as a secondary effect of CDW rather than the proof of the relevance of electron-phonon coupling for the electron subsystem\cite{Reznik}.

Beside the phonon scenario, the spectral anomalies observed in the ARPES measurements have also been interpreted in terms of quasiparticle scattering from the neutron resonance mode. The problem with such a scenario is that the spectral weight of the neutron resonance mode is not only very small\cite{Kee}, but also strongly temperature and doping dependent. More generally, no matter what kind of mode is assumed, a common problem with the mode coupling scenario is that the observed spectral anomalies are always accompanied by rather broad spectral continuum at higher energy. This would imply the importance of higher order scattering processes in the mode coupling picture, which has never been seriously considered. In a recent investigation of this issue, we point out that the peak-dip-hump structure around the anti-nodal region should be interpreted in terms of the coupling between the BCS quasiparticle and the ubiquitous diffusive antiferromagnetic spin fluctuation in the system, whose existence has been extensively verified by RIXS measurement on the cuprate superconductors\cite{Tacon,Dean}. In particular, we find that the coupling between the quasiparticle and the diffusive antiferromagnetic spin fluctuation is responsible for the exceptional flatness of the anti-nodal dispersion\cite{Dessau}. According to this scenario, far from being a standard d-wave BCS Bogliubov quasiparticle, the sharp coherence peak found in the anti-nodal spectrum should be viewed as a composite object made up of the Bogliubov quasiparticle and the intense antiferromagnetic spin fluctuation dressing cloud.

The strong non-BCS character of the anti-nodal quasiparticle excitation indicates that it would also experience phonon scattering in a way very different from that for the bare BCS Bogliubov quasiparticle. Such a difference is caused by the vertex correction of the electron-phonon coupling from the antiferromagnetic spin fluctuation, as is required by the Ward identity. In this paper, we show that the coupling to the $B_{1g}$ buckling mode suffers from a destructive interference effect from such a vertex correction. In the real space scenario, it simply amounts to the fact that the electron hopping between nearest neighboring sites is suppressed by the antiferromagnetic spin correlation between the electrons. Thus, unlike weakly correlated system in which the quasiparticle scattering rate from different scattering channels simply add up, strong vertex correction from the antiferromagnetic spin fluctuation tends to suppress the coupling between the electron and the $B_{1g}$ phonon in the cuprate superconductors. Such an effect is expected to be enhanced as we move closer to the Mott insulating phase with underdoping. More generally, we argue that with the proximity of the Mott insulating limit, the electron-phonon coupling in both the diagonal and the off-diagonal channel should diminish as a result of electron correlation effect. We thus believe that the absence of the phonon signature in the dc resistivity should be attributed to the vertex correction effect from electron correlation effect.

We assume that the electron dispersion of the system is given by the following tight binding model
\begin{eqnarray}
H=&-t&\sum_{i,\bm{\delta},\sigma}( c^{\dagger}_{i,\sigma}c_{i+\bm{\delta},\sigma}+h.c.)\nonumber\\
&-t'&\sum_{i,\bm{\delta'},\sigma}( c^{\dagger}_{i,\sigma}c_{i+\bm{\delta'},\sigma}+h.c.)-\mu\sum_{i,\sigma} c^{\dagger}_{i,\sigma}c_{i,\sigma}
\end{eqnarray}
Here $\bm{\delta}=\mathbf{x,y}$ denotes the lattice vector connecting the nearest-neighboring(NN) sites on the square lattice, $\bm{\delta'}=\mathbf{x\pm y}$ denotes the lattice vector connecting the next-nearest-neighboring(NNN) sites on the square lattice, $t$ and $t'$ are the NN and NNN hopping integral, $\mu$ is the chemical potential. To simulate the electron correlation effect, we assume that the electron is coupled to a fluctuating background of antiferromagnetic correlated local moments as in the spin-fermion model\cite{Chubukov}. The coupling is given by
\begin{equation}
H_{sf}=g\sum_{i}\vec{S}_{i}\cdot \vec{s}_{i}
\end{equation} 
in which $\vec{S}_{i}$ denotes the fluctuating local moment at site $i$ and
\begin{equation} 
\vec{s}_{i}=\frac{1}{2}\sum_{\alpha,\beta}c^{\dagger}_{i,\alpha}\bm{\sigma}_{\alpha,\beta}c_{i,\beta}
\end{equation}
is the electron spin density at site $i$. We assume that the fluctuation of the local moment is described by a Gaussian action of the form
\begin{equation}
S_{eff}=\frac{1}{\beta} \sum_{\mathbf{q},i\omega_{n}}\chi^{-1}(\mathbf{q},i\omega_{n})\vec{S}(\mathbf{q},i\omega_{n})\cdot \vec{S}(-\mathbf{q},-i\omega_{n})
\end{equation}
in which $\chi(\mathbf{q},i\omega_{n})$ denotes the generalized susceptibility of the local moment. Here we adopt the widely used Millis-Monien-Pines(MMP) form for $\chi(\mathbf{q},i\omega_{n})$, which reads\cite{MMP}
\begin{equation}
\chi(\mathbf{q},i\omega_{n})=\frac{\chi_{0}}{1+\xi^{2}(\mathbf{q}-\mathbf{Q})^{2}+\frac{|\omega_{n}|}{\omega_{sf}}}
\end{equation}
Here $\omega_{sf}$ denotes the characteristic frequency of the Landau damped local moment fluctuation, $\xi$ denotes the correlation length of the local moment, $\chi_{0}$ is the static spin susceptibility of the local moment at the antiferromagnetic wave vector. As will be clear below, the overall scale for the coupling between the electron and the local moment is set by the product $g^{2}\chi_{0}$.

We assume that the electron-phonon coupling with the $B_{1g}$ buckling mode take the linear Su-Schrieffer-Hegger(SSH) form\cite{Nagaosa,Shen}, which reads
\begin{equation}
H_{SSH}=\lambda\sum_{i,\bm{\delta},\sigma}u_{i,\bm{\delta}}( c^{\dagger}_{i,\sigma}c_{i+\bm{\delta},\sigma}+h.c.)
\end{equation} 
here $u_{i,\bm{\delta}}$ is the phonon coordinate of the $B_{1g}$ mode on the bond connecting site $i$ and $i+\bm{\delta}$. $\lambda$ is the bare electron-phonon coupling strength. Note that such a linear coupling is only permitted when the position of the oxygen ion is away from the inversion center and is thus not universally applicable for the cuprate superconductors. For systems that lack such a linear coupling we expect the electron-phonon coupling effect to be more strongly reduced. For simplicity, we neglect the dispersion of the $B_{1g}$ buckling mode and assume it to fluctuate locally with a frequency of $\Omega$. 

A crucial feature of the SSH coupling can be seen more clearly if we go to the Fourier space, in which $H_{SSH}$ takes the form of 
\begin{equation}
H_{SSH}=\frac{1}{\sqrt{N}}\sum_{\mathbf{k,q},\sigma\bm{\delta}}F_{\bm{\delta}}(\mathbf{k,q})u_{\bm{\delta}}(\mathbf{q})c^{\dagger}_{\mathbf{k-q},\sigma}c_{\mathbf{k},\sigma}
\end{equation}
here $N$ denotes the number of lattice site. $u_{\bm{\delta}}(\mathbf{q})$ and $c_{\mathbf{k},\sigma}$ are the Fourier component of the phonon operator and the electron operator and are given by
\begin{eqnarray}
u_{\bm{\delta}}(\mathbf{q})&=&\frac{1}{\sqrt{N}}\sum_{i}u_{i,\bm{\delta}}e^{i\mathbf{q}\cdot\mathbf{r}_{i}}\nonumber\\
c_{\mathbf{k},\sigma}&=&\frac{1}{\sqrt{N}}\sum_{i}c_{i,\sigma}e^{-i\mathbf{q}\cdot\mathbf{r}_{i}}
\end{eqnarray}
$F_{\bm{\delta}}(\mathbf{k,q})$ is the bare electron-phonon vertex in the momentum space and is given by
\begin{equation}
F_{\bm{\delta}}(\mathbf{k,q})=\lambda[e^{i\mathbf{k}\cdot \bm{\delta}}+e^{-i\mathbf{(k-q)}\cdot \bm{\delta}}]
\end{equation}
Since $\bm{\delta}$ connects sites in different sublattice of the square lattice, we have
\begin{equation}
F_{\bm{\delta}}(\mathbf{k,q})=-F_{\bm{\delta}}(\mathbf{k+Q,q})
\end{equation}
for any $\mathbf{k}$ and $\mathbf{q}$. Here $\mathbf{Q}=(\pi,\pi)$ is the antiferromagnetic wave vector. As we will see below, it is this feature of the electron-phonon coupling that is responsible for the destructive interference effect in the vertex correction by the antiferromagnetic spin fluctuation.
 
  \begin{figure}
\includegraphics[width=8cm]{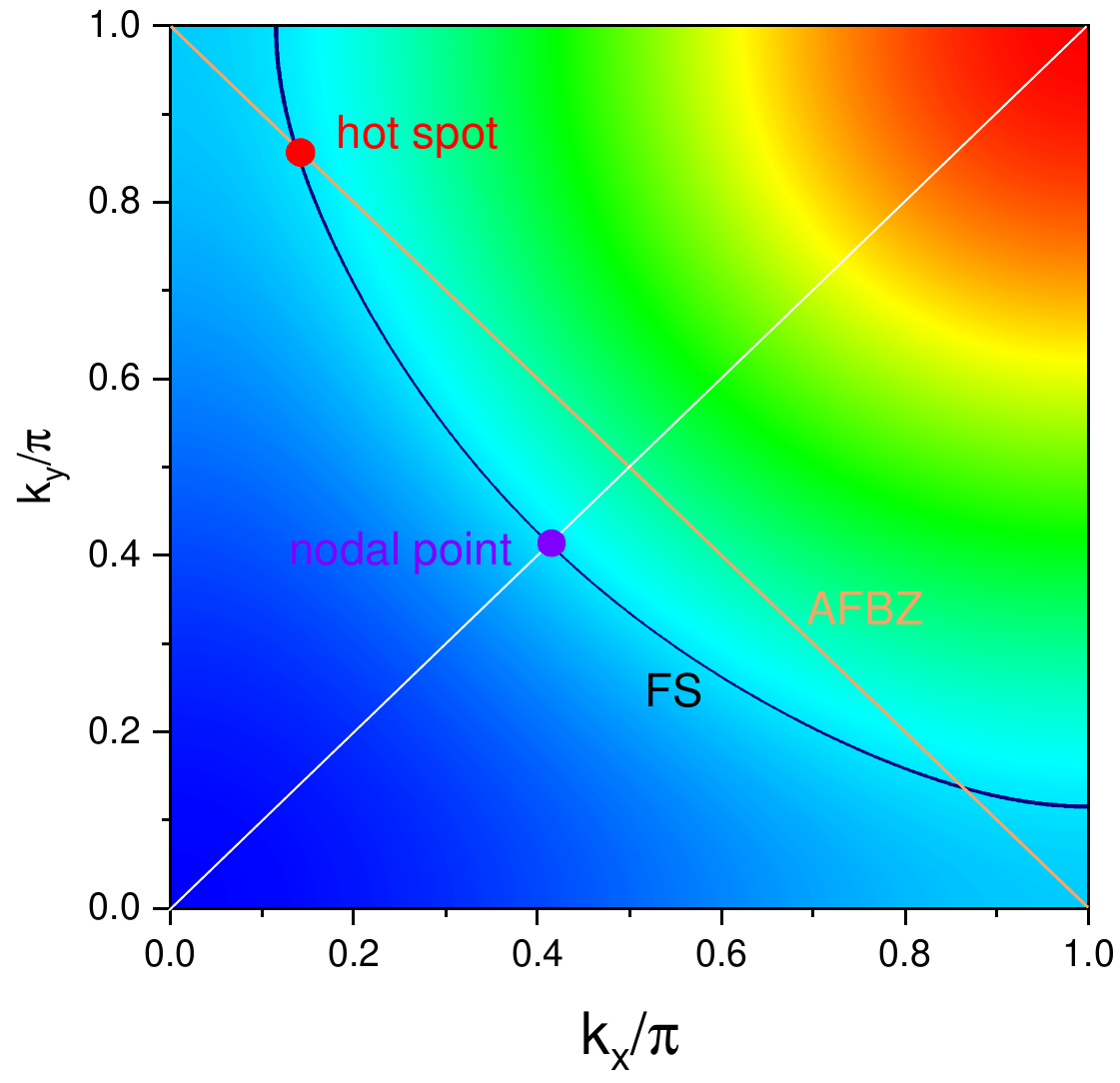}
\caption{The nodal point and the hot spot on the fermi surface as denoted by the black curve for a system with $t'=-0.3t$ and $\mu=-t$. Shown here is the first quadrant of the Brillouin zone. The orange straight line from $(\pi,0)$ to $(0,\pi)$ denotes the boundary of the antiferromagnetic Brillouin zone. It crosses the fermi surface at the hot spots, where the effect of the scattering from antiferromagnetic spin fluctuation is the strongest.}
\end{figure}   
 
In this study, we set $t=250\ meV$ and use it as the unit of energy. This is also the most commonly adopted value in the literature for the cuprate superconductors. We will set $t'=-0.3t$ and $\mu=-t$. This corresponds to a system with hole doping level of about $x=15\%$ and a fermi surface as depicted in Fig.1. We will set $\xi=3$ and $\omega_{sf}=0.04t$ for the spin fluctuation spectrum. These values are typical for optimally doped cuprate superconductors\cite{Zha,Zhu}. $\chi_{0}$ can be estimated from the local spin sum rule, which requires
\begin{equation}
\langle \mathbf{S}^{2}_{i}\rangle=\frac{1}{2\pi N}\sum_{\mathbf{q}}\int_{0}^{\omega_{c}}d\omega \coth(\frac{\beta\omega}{2})R(\mathbf{q},\omega)=\frac{3}{4}(1-x)
\end{equation}
in which
\begin{equation}
R(\mathbf{q},\omega)=\chi_{0}\frac{2\omega/\omega_{sf}}{(1+\xi^{2}(\mathbf{q-Q})^{2})^{2}+(\omega/\omega_{sf})^{2}}
\end{equation}
is the spectral weight of the local moment fluctuation. Here $\omega_{c}$ is the high energy cutoff of the spin fluctuation spectrum and we set it to be $\omega_{c}=2t=500 \ meV$\cite{Tacon}. Completing the above integration we find that $\chi_{0}=400\ eV^{-1}=100/t$. Such a value is very consistent with that estimated from NMR data\cite{Zha}. The coupling strength $g$ between the spin fluctuation and the electron is more difficult to estimate and will treated as a tunable parameter. In most of our computation we will set $g=t$. We set $\Omega=0.2t$ and adopt a value of $\lambda=0.2 t$ for the electron-phonon coupling. Such a value is consistent with previous experimental and theoretical estimates\cite{Nagaosa,Shen,Peng}.

 \begin{figure}
\includegraphics[width=12cm]{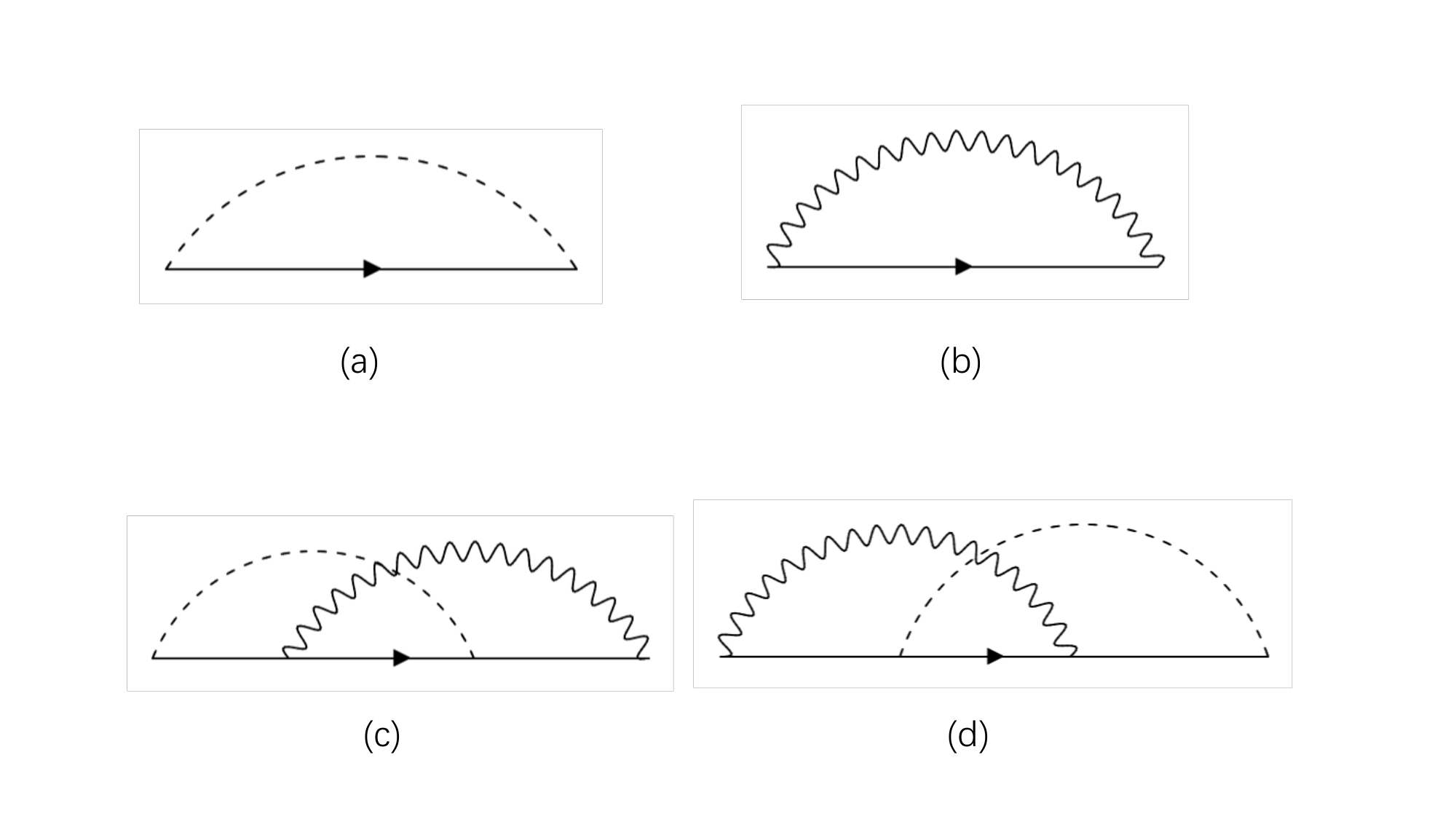}
\caption{The electron self energy to the lowest order in the coupling strength $\lambda$ and $g$. (a)The lowest order self energy caused by the antiferromagnetic spin fluctuation. Here the solid line denotes the electron Green's function, the dashed line denotes the propagator of the spin fluctuation. (b) The lowest order self energy caused by the $B_{1g}$ buckling phonon. Here the wavy line denotes the phonon propagator. (c) The lowest order intertwining term in the self energy expansion between the antiferromagnetic spin fluctuation and $B_{1g}$ phonon. The intertwining term in (c) can be interpreted either as the first order vertex correction to the electron-phonon vertex in (b) by the antiferromagnetic spin fluctuation, or, as the first order vertex correction to the spin-fermion coupling in (a) by the $B_{1g}$ phonon.}
\end{figure}  
 
 At the lowest perturbative order in the coupling strength $g$ and $\lambda$, the self energy of the quasiparticle is given by the Feynman diagrams shown in Fig.2. Here Fig.2a denotes the self energy correction caused by the scattering from the antiferromagnetic spin fluctuation. Fig.2b denotes the self energy correction caused by the coupling to the $B_{1g}$ phonon. Fig.2c denotes the intertwining contribution between the antiferromagnetic spin fluctuation and the $B_{1g}$ phonon. As we will show below, such an intertwinement would result in the destructive interference effect in the effective electron-phonon coupling strength in the system. We note that the intertwining term in Fig.2c can be interpreted either as the first order vertex correction to the electron-phonon vertex in Fig.2b by the antiferromagnetic spin fluctuation, or, as the first order vertex correction to the spin-fermion coupling in Fig.2a by the $B_{1g}$ phonon.

Before delving into the perturbative analysis of the electron self energy we first consider the extreme case in which the antiferromagnetic spin fluctuation freezes into static and long-ranged antiferromagnetic order. This corresponds to the limit of $\xi\rightarrow \infty$ and $\omega_{sf}\rightarrow 0$ in Eq.6. The electron Hamiltonian in this limit reads
\begin{eqnarray}
H_{MF}=\sum_{\mathbf{k\in AFBZ},\ \sigma }[\ \epsilon_{\mathbf{k}}c^{\dagger}_{\mathbf{k},\sigma}c_{\mathbf{k},\sigma}+\epsilon_{\mathbf{k+Q}}c^{\dagger}_{\mathbf{k+Q},\sigma}c_{\mathbf{k+Q},\sigma}\ ]
-\frac{gm}{2}\sum_{\mathbf{k\in AFBZ},\ \sigma}\sigma\ [c^{\dagger}_{\mathbf{k+Q},\sigma}c_{\mathbf{k},\sigma}+c^{\dagger}_{\mathbf{k},\sigma}c_{\mathbf{k+Q},\sigma}\ ]\nonumber\\
\end{eqnarray}
Here $\mathbf{AFBZ}$ denotes the antiferromagnetic Brillouin zone(whose boundary is depicted as the orange line in Fig.1), $m=|\langle \vec{S}_{i}\rangle|$ is the length of the antiferromagnetic ordered moment. The electron dispersion is given by
\begin{equation}
\epsilon_{\mathbf{k}}=-2t(\cos k_{x}+\cos k_{y})-4t'\cos k_{x} \cos k_{y}-\mu
\end{equation}
We are now interested in how the electron-phonon scattering matrix element is renormalized in the antiferromagnetic ordered background. For this purpose, we diagonalize Eq.14 by the unitary transformation
\begin{equation}
\Psi_{\mathbf{k}}=\left(\begin{array}{c}c_{\mathbf{k},\sigma} \\c_{\mathbf{k+Q},\sigma}\end{array}\right)=U_{\mathbf{k}}\bm{\gamma}_{\mathbf{k}}=\left(\begin{array}{cc}u_{\mathbf{k}} & -v_{\mathbf{k}} \\v_{\mathbf{k}} & u_{\mathbf{k}}\end{array}\right)\left(\begin{array}{c}\gamma_{\mathbf{k,+},\sigma} \\\gamma_{\mathbf{k,-},\sigma}\end{array}\right)
\end{equation}
in which $\gamma_{\mathbf{k,\pm},\sigma}$ denotes the operator for the quasiparticle in the upper and the lower SDW band whose energy is given by 
\begin{equation}
E_{\mathbf{k,\pm}}=\frac{( \epsilon_{\mathbf{k}}+ \epsilon_{\mathbf{k+Q}})\pm\sqrt{( \epsilon_{\mathbf{k}}- \epsilon_{\mathbf{k+Q}})^{2}+(gm)^{2}}}{2}
\end{equation}
The electron-phonon coupling Hamiltonian Eq.8 can be rewritten in terms of the $\bm{\gamma}$ operators as follows
\begin{eqnarray}
H_{SSH}&=&\frac{1}{\sqrt{N}}\sum_{\mathbf{k,q},\sigma,\bm{\delta}}F_{\bm{\delta}}(\mathbf{k,q})u_{\bm{\delta}}(\mathbf{q})c^{\dagger}_{\mathbf{k+q},\sigma}c_{\mathbf{k},\sigma}\nonumber\\
&=&\frac{1}{\sqrt{N}}\sum_{\mathbf{q},\bm{\delta}}u_{\bm{\delta}}(\mathbf{q})\sum_{\mathbf{k\in AFBZ},\sigma}F_{\bm{\delta}}(\mathbf{k,q})\Psi^{\dagger}_{\mathbf{k+q}}\tau_{3}\Psi_{\mathbf{k}}\nonumber\\
&=&\frac{1}{\sqrt{N}}\sum_{\mathbf{q},\bm{\delta}}u_{\bm{\delta}}(\mathbf{q})\sum_{\mathbf{k\in AFBZ},\sigma}F_{\bm{\delta}}(\mathbf{k,q})\bm{\gamma}^{\dagger}_{\mathbf{k+q}}\mathbf{M_{k,q}}\bm{\gamma}_{\mathbf{k}}\nonumber\\
\end{eqnarray}
in which $\tau_{3}=\left(\begin{array}{cc}1 & 0 \\0 & -1\end{array}\right)$ and $\mathbf{M_{k,q}}=U^{\dagger}_{\mathbf{k+q}}\tau_{3}U_{\mathbf{k}}$.
We note that to obtain the second and the third line, we have used the property $F_{\bm{\delta}}(\mathbf{k,q})=-F_{\bm{\delta}}(\mathbf{k+Q,q})$. We have also extended the definition of $\Psi_{\mathbf{k}}$, $\bm{\gamma}_{\mathbf{k}}$ and $U_{\mathbf{k}}$ from the $\mathbf{AFBZ}$ to the full Brillouin zone of the square lattice in such a way as
\begin{eqnarray}
\Psi_{\mathbf{k+Q}}&=&\tau_{1}\Psi_{\mathbf{k}}\nonumber\\
\bm{\gamma}_{\mathbf{k+Q}}&=&\bm{\gamma}_{\mathbf{k}}\nonumber\\
U_{\mathbf{k+Q}}&=&\tau_{1}U_{\mathbf{k}}
\end{eqnarray}
in which $\tau_{1}=\left(\begin{array}{cc}0 & 1 \\1 & 0\end{array}\right)$. The matrix elements of $\mathbf{M_{k,q}}$ are given as follows. For $\mathbf{k,k+q\in AFBZ}$, we have
\begin{equation}
\mathbf{M_{k,q}}=\left(\begin{array}{cc}M_{11} & -M_{12} \\-M_{12} & -M_{11}\end{array}\right)
\end{equation}
in which 
\begin{eqnarray}
M_{11}&=&u_{\mathbf{k+q}}u_{\mathbf{k}}-v_{\mathbf{k+q}}v_{\mathbf{k}}\nonumber\\
M_{12}&=&u_{\mathbf{k+q}}v_{\mathbf{k}}+v_{\mathbf{k+q}}u_{\mathbf{k}}
\end{eqnarray}
While for $\mathbf{k\in AFBZ}$ and $\mathbf{k+q\notin AFBZ}$, we have
\begin{equation}
\mathbf{M_{k,q}}=\left(\begin{array}{cc}\tilde{M}_{11} & \tilde{M}_{12} \\-\tilde{M}_{12} & \tilde{M}_{11}\end{array}\right)
\end{equation}
in which 
\begin{eqnarray}
\tilde{M}_{11}&=&v_{\mathbf{k+q+Q}}u_{\mathbf{k}}-u_{\mathbf{k+q+Q}}v_{\mathbf{k}}\nonumber\\
\tilde{M}_{12}&=&u_{\mathbf{k+q+Q}}u_{\mathbf{k}}+v_{\mathbf{k+q+Q}}u_{\mathbf{k}}
\end{eqnarray}
Note that if $\mathbf{k+q\notin AFBZ}$, then $\mathbf{k+q+Q\in AFBZ}$.  Let us focus on the intra-band matrix element of the electron-phonon coupling.  From Eq.21 and Eq.23, it is clear that both $M_{11}$ and $\tilde{M}_{11}$ are suppressed in the antiferromagnetic ordered background. In particular, the intra-band matrix element of the electron-phonon coupling becomes exactly zero at the hot spots. Such a suppression is easy to understand from the real space perspective, since the nearest neighboring hopping of the electron that the $B_{1g}$ mode couple to is suppressed by the antiferromagnetic order.

We now perform the full perturbative analysis of the electron self energy with more realistic parameters. The self energy caused by the antiferromagnetic spin fluctuation as shown in Fig.2a is given by
\begin{equation}
\Sigma^{AF}(\mathbf{k},i\nu_{n})=-\frac{3g^{2}}{\beta N}\sum_{\mathbf{q},i\omega_{n}}\chi(\mathbf{q},i\omega_{n})G(\mathbf{k-q},i\nu_{n}-i\omega_{n})
\end{equation}
Here the prefactor $3$ is due to the fact that there are three independent components in the local moment fluctuation. Introducing the spectral representation of $\chi(\mathbf{q},i\omega_{n})$
\begin{equation}
\chi(\mathbf{q},i\omega_{n})=\frac{1}{2\pi}\int_{0}^{\infty}\frac{-2\omega}{\omega_{n}^{2}+\omega^{2}} R(\mathbf{q},\omega)d\omega
\end{equation} 
and completing the frequency sum we have
\begin{eqnarray}
\Sigma^{AF}(\mathbf{k},i\nu_{n})=\frac{3g^{2}}{2\pi N}\sum_{\mathbf{q}}\int_{0}^{\infty}d\omega R(\mathbf{q},\omega)
\times\left\{\frac{1-n_{F}(\epsilon_{\mathbf{k-q}})+n_{B}(\omega)}{i\nu_{n}-\omega-\epsilon_{\mathbf{k-q}}} 
+ \frac{n_{F}(\epsilon_{\mathbf{k-q}})+n_{B}(\omega)}{i\nu_{n}+\omega-\epsilon_{\mathbf{k-q}}} \right\}\nonumber\\
\end{eqnarray}
Similarly, the self energy caused by electron-phonon coupling as shown in Fig.2b is given by
\begin{eqnarray}
\Sigma^{ph}(\mathbf{k},i\nu_{n})=-\frac{1}{\beta N}\sum_{\mathbf{q},i\omega_{n},\bm{\delta}}|F_{\bm{\delta}}(\mathbf{k,q})|^{2}D_{\bm{\delta}}(\mathbf{q},i\omega_{n})G(\mathbf{k-q},i\nu_{n}-i\omega_{n})
\end{eqnarray}
 The phonon propagator is given by
\begin{equation}
D_{\bm{\delta}}(\mathbf{q},i\omega_{n})=\frac{-2\Omega}{\omega_{n}^{2}+\Omega^{2}}
\end{equation}
The phonon part of the electron self energy is then given by
 \begin{eqnarray}
\Sigma^{ph}(\mathbf{k},i\nu_{n})=\frac{1}{N}\sum_{\mathbf{q},\bm{\delta}}|F_{\bm{\delta}}(\mathbf{k,q})|^{2}
\times\left\{\frac{1-n_{F}(\epsilon_{\mathbf{k-q}})+n_{B}(\Omega)}{i\nu_{n}-\Omega-\epsilon_{\mathbf{k-q}}} 
+ \frac{n_{F}(\epsilon_{\mathbf{k-q}})+n_{B}(\Omega)}{i\nu_{n}+\Omega-\epsilon_{\mathbf{k-q}}} \right\}\nonumber\\
\end{eqnarray}

 \begin{figure}
\includegraphics[width=9cm]{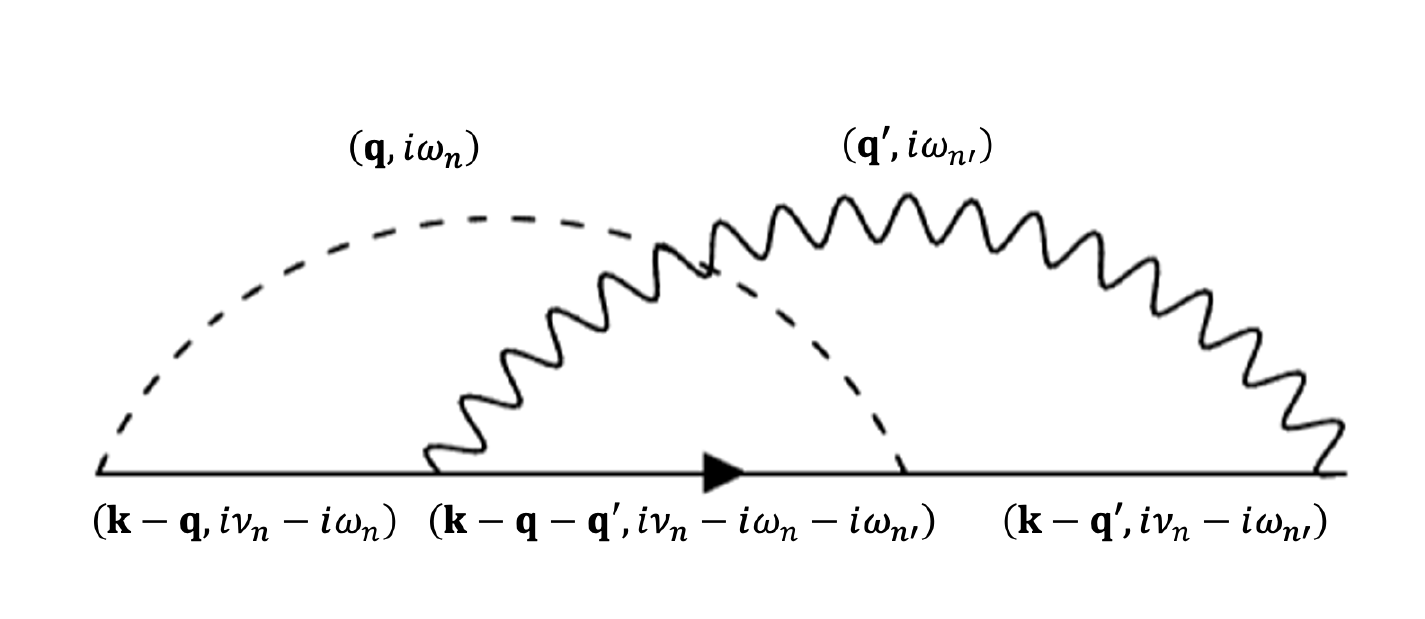}
\caption{The electron self energy contributed by the intertwined scattering from the $B_{1g}$ phonon and the antiferromagnetic spin fluctuation. Here the solid lines denote the electron Green's function, the dashed line denotes the propagator of the spin fluctuation, and the wavy line denotes the phonon propagator. Note that such an intertwined contribution can be interpreted either as the vertex correction to $\Sigma^{ph}$ from the antiferromagnetic spin fluctuation, or, as the vertex correction to $\Sigma^{AF}$ from the phonon.}
\end{figure}

The intertwined terms between the $B_{1g}$ phonon and the antiferromagnetic spin fluctuation has a more complicated form. In Fig.3, we illustrate the structure of one of such term. Following the Feynman rule, we get
\begin{eqnarray}
\Sigma^{ver}(\mathbf{k},i\nu_{n})=\frac{6 g^{2}}{\beta^{2} N^{2}}\sum_{\mathbf{q,q'},\bm{\delta},i\omega_{n},i\omega_{n'}}&&\mathbf{Re}[F_{\bm{\delta}}(\mathbf{k-q},\mathbf{q'})F^{*}_{\bm{\delta}}(\mathbf{k},\mathbf{q'})]\ 
\chi(\mathbf{q},i\omega_{n}) D_{\bm{\delta}}(\mathbf{q'},i\omega_{n'})
G(\mathbf{k-q},i\nu_{n}-i\omega_{n})\nonumber\\
&\times& G(\mathbf{k-q'},i\nu_{n}-i\omega_{n'})G(\mathbf{k-q-q'},i\nu_{n}-i\omega_{n}-i\omega_{n'})\nonumber\\
\end{eqnarray} 
Here $\mathbf{q}$ and $\mathbf{q'}$ denote the wave vector of the antiferromagnetic spin fluctuation and the $B_{1g}$ phonon. $i\omega_{n}$ and $i\omega_{n'}$ denote their respective Mastubara frequencies. We note that such an intertwining term can be interpreted either as the vertex correction to the electron-phonon coupling by the antiferromagnetic spin fluctuation, or, as the vertex correction to the spin-fermion coupling by the $B_{1g}$ phonon. Here we will focus on the former aspect of the intertwining term since what concerns us here is the electron scattering at a temperature scale of the order of the phonon frequency $\Omega$, which is substantially higher than $\omega_{sf}$.

 Inserting the spectral representation of $\chi(\mathbf{q},i\omega_{n})$ and completing the sum over the Mastubara frequency $i\omega_{n}$ and $i\omega_{n'}$, we arrive at the following expression for the intertwining term
\begin{eqnarray}
\Sigma^{ver}(\mathbf{k},i\nu_{n})=\frac{3g^{2}}{\pi N^{2}}\sum_{\mathbf{q,q'},\bm{\delta}}\mathbf{Re}[F_{\bm{\delta}}(\mathbf{k-q},\mathbf{q'})F^{*}_{\bm{\delta}}(\mathbf{k},\mathbf{q'})]
\int_{0}^{\infty}d\omega R(\mathbf{q},\omega)\sum_{\sigma,\sigma'=\pm1}\sigma\sigma'A[\mathbf{k,q,q'},i\nu_{n},\sigma \omega,\sigma' \Omega]\nonumber\\
\end{eqnarray} 
in which $A$ has the lengthy expression of
\begin{eqnarray}
&&A[\mathbf{k,q,q'},i\nu_{n}, \omega,\Omega]\nonumber\\
&&=n_{B}(\Omega)n_{B}(\omega)\times  \frac{1}{i\nu_{n}-\omega-\epsilon_{\mathbf{k-q}}}\times\frac{1}{i\nu_{n}-\Omega-\epsilon_{\mathbf{k-q'}}}\times\frac{1}{i\nu_{n}-\omega-\Omega-\epsilon_{\mathbf{k-q-q'}}}\nonumber\\
&&-n_{F}(-\epsilon_{\mathbf{k-q}})n_{F}(-\epsilon_{\mathbf{k-q'}})\times  \frac{1}{i\nu_{n}-\omega-\epsilon_{\mathbf{k-q}}}\times\frac{1}{i\nu_{n}-\Omega-\epsilon_{\mathbf{k-q'}}}\times\frac{1}{i\nu_{n}-\epsilon_{\mathbf{k-q}}-\epsilon_{\mathbf{k-q'}}+\epsilon_{\mathbf{k-q-q'}}}\nonumber\\
&&+\left[\frac{n_{B}(\Omega)n_{F}(-\epsilon_{\mathbf{k-q}})}{\Delta_{1}}+\frac{n_{B}(\omega)n_{F}(-\epsilon_{\mathbf{k-q'}})}{\Delta_{2}}\right]\times\frac{1}{i\nu_{n}-\omega-\epsilon_{\mathbf{k-q}}}\times\frac{1}{i\nu_{n}-\Omega-\epsilon_{\mathbf{k-q'}}}\nonumber\\
&&-\frac{1}{2}\frac{\coth(\frac{\beta\omega}{2})n_{F}(-\epsilon_{\mathbf{k-q-q'}})}{\Delta_{2}}\times \frac{1}{i\nu_{n}-\omega-\epsilon_{\mathbf{k-q}}}\times\frac{1}{i\nu_{n}-\omega-\Omega-\epsilon_{\mathbf{k-q-q'}}}\nonumber\\
&&-\frac{1}{2}\frac{\coth(\frac{\beta\Omega}{2})n_{F}(-\epsilon_{\mathbf{k-q-q'}})}{\Delta_{1}}\times \frac{1}{i\nu_{n}-\Omega-\epsilon_{\mathbf{k-q'}}}\times\frac{1}{i\nu_{n}-\omega-\Omega-\epsilon_{\mathbf{k-q-q'}}}\nonumber\\
&&+\frac{1}{2}\frac{\tanh(\frac{\beta\epsilon_{\mathbf{k-q}}}{2})n_{F}(-\epsilon_{\mathbf{k-q-q'}})}{\Delta_{1}}\times \frac{1}{i\nu_{n}-\omega-\epsilon_{\mathbf{k-q}}}\times\frac{1}{i\nu_{n}-\epsilon_{\mathbf{k-q}}-\epsilon_{\mathbf{k-q'}}+\epsilon_{\mathbf{k-q-q'}}}\nonumber\\
&&+\frac{1}{2}\frac{\tanh(\frac{\beta\epsilon_{\mathbf{k-q'}}}{2})n_{F}(-\epsilon_{\mathbf{k-q-q'}})}{\delta_{2}}\times \frac{1}{i\nu_{n}-\Omega-\epsilon_{\mathbf{k-q'}}}\times\frac{1}{i\nu_{n}-\epsilon_{\mathbf{k-q}}-\epsilon_{\mathbf{k-q'}}+\epsilon_{\mathbf{k-q-q'}}}\nonumber\\
\end{eqnarray}
Here
\begin{eqnarray}
\Delta_{1}=\epsilon_{\mathbf{k-q}}-\epsilon_{\mathbf{k-q-q'}}-\Omega\nonumber\\
\Delta_{2}=\epsilon_{\mathbf{k-q'}}-\epsilon_{\mathbf{k-q-q'}}-\omega
\end{eqnarray}

We now derive the retarded self energy of electron by performing the analytic continuation $i\nu_{n}\rightarrow \nu+i0^{+}$ in $\Sigma(\mathbf{k},i\nu_{n})$. The imaginary part of the retarded self energy due to the antiferromagnetic spin fluctuation alone is given by
\begin{eqnarray}
-\mathbf{Im}\Sigma^{AF}(\mathbf{k},\nu+i0^{0})=\frac{3g^{2}}{2 N}\sum_{\mathbf{q}}R(\mathbf{q},|\nu-\epsilon_{\mathbf{k-q}}|)[n_{F}(\mathbf{sgn}(\epsilon_{\mathbf{k-q}}-\nu)\epsilon_{\mathbf{k-q}})+n_{B}(|\nu-\epsilon_{\mathbf{k-q}}|)]
\end{eqnarray} 
Similarly, the self energy due to the $B_{1g}$ phonon is given by
\begin{eqnarray}
-\mathbf{Im}\Sigma^{ph}(\mathbf{k},\nu+i0^{+})=\frac{\pi}{N}\sum_{\mathbf{q},\bm{\delta}}|F_{\bm{\delta}}(\mathbf{k,q})|^{2}
&&[(1-n_{F}(\epsilon_{\mathbf{k-q}})+n_{B}(\Omega))\delta(\nu-\Omega-\epsilon_{\mathbf{k-q}})\nonumber\\
&&+(n_{F}(\epsilon_{\mathbf{k-q}})+n_{B}(\Omega))\delta(\nu+\Omega-\epsilon_{\mathbf{k-q}})]\nonumber\\
\end{eqnarray} 
The intertwining term is much more complicated and is given by 
\begin{eqnarray}
-\mathbf{Im}\Sigma^{ver}(\mathbf{k},\nu+i0^{+})=-\frac{3g^{2}}{\pi N^{2}}\sum_{\mathbf{q,q'},\bm{\delta}}\mathbf{Re}[F_{\bm{\delta}}(\mathbf{k-q},\mathbf{q'})F^{*}_{\bm{\delta}}(\mathbf{k},\mathbf{q'})]
\int_{0}^{\infty}d\omega R(\mathbf{q},\omega)\sum_{\sigma,\sigma'=\pm1}(\sigma\sigma') \mathbf{Im} A[\mathbf{k,q,q'},\nu+i0^{+},\sigma \omega,\sigma' \Omega]\nonumber\\
\end{eqnarray} 
Here 
\begin{eqnarray}
&&-\mathbf{Im}\ A[\mathbf{k,q,q'},\nu+i0^{+}, \omega,\Omega]\nonumber\\
&&=\pi\frac{n_{B}(\omega)+n_{F}(-\epsilon_{\mathbf{k-q}})}{\Delta_{1}-\Delta_{2}}  \times [\frac{n_{B}(\Omega)+n_{F}(-\epsilon_{\mathbf{k-q-q'}})}{\Delta_{1}}+ \frac{n_{F}(-\epsilon_{\mathbf{k-q'}})-n_{F}(-\epsilon_{\mathbf{k-q-q'}})}{\Delta_{2}}   ]\delta(\nu-\omega-\epsilon_{\mathbf{k-q}})\nonumber\\
&&+\pi\frac{n_{B}(\Omega)+n_{F}(-\epsilon_{\mathbf{k-q'}})}{\Delta_{2}-\Delta_{1}}  \times [\frac{n_{B}(\omega)+n_{F}(-\epsilon_{\mathbf{k-q-q'}})}{\Delta_{2}}+ \frac{n_{F}(-\epsilon_{\mathbf{k-q}})-n_{F}(-\epsilon_{\mathbf{k-q-q'}})}{\Delta_{1}}   ]\delta(\nu-\Omega-\epsilon_{\mathbf{k-q'}})\nonumber\\
&&+\pi\frac{1}{\Delta_{1}\Delta_{2}}  \times [n_{B}(\omega)n_{B}(\Omega)+n_{F}(-\epsilon_{\mathbf{k-q-q'}})(1+n_{B}(\omega)+n_{B}(\Omega))]\delta(\nu-\omega-\Omega-\epsilon_{\mathbf{k-q-q'}})\nonumber\\
&&+\pi\frac{1}{\Delta_{1}\Delta_{2}}  \times [n_{F}(-\epsilon_{\mathbf{k-q-q'}})(1-n_{F}(\epsilon_{\mathbf{k-q}})-n_{F}(\epsilon_{\mathbf{k-q'}}))-n_{F}(-\epsilon_{\mathbf{k-q}})n_{F}(-\epsilon_{\mathbf{k-q'}})]\delta(\nu-\epsilon_{\mathbf{k-q}}-\epsilon_{\mathbf{k-q'}}+\epsilon_{\mathbf{k-q-q'}})\nonumber\\
\end{eqnarray}

 \begin{figure}
\includegraphics[width=8cm]{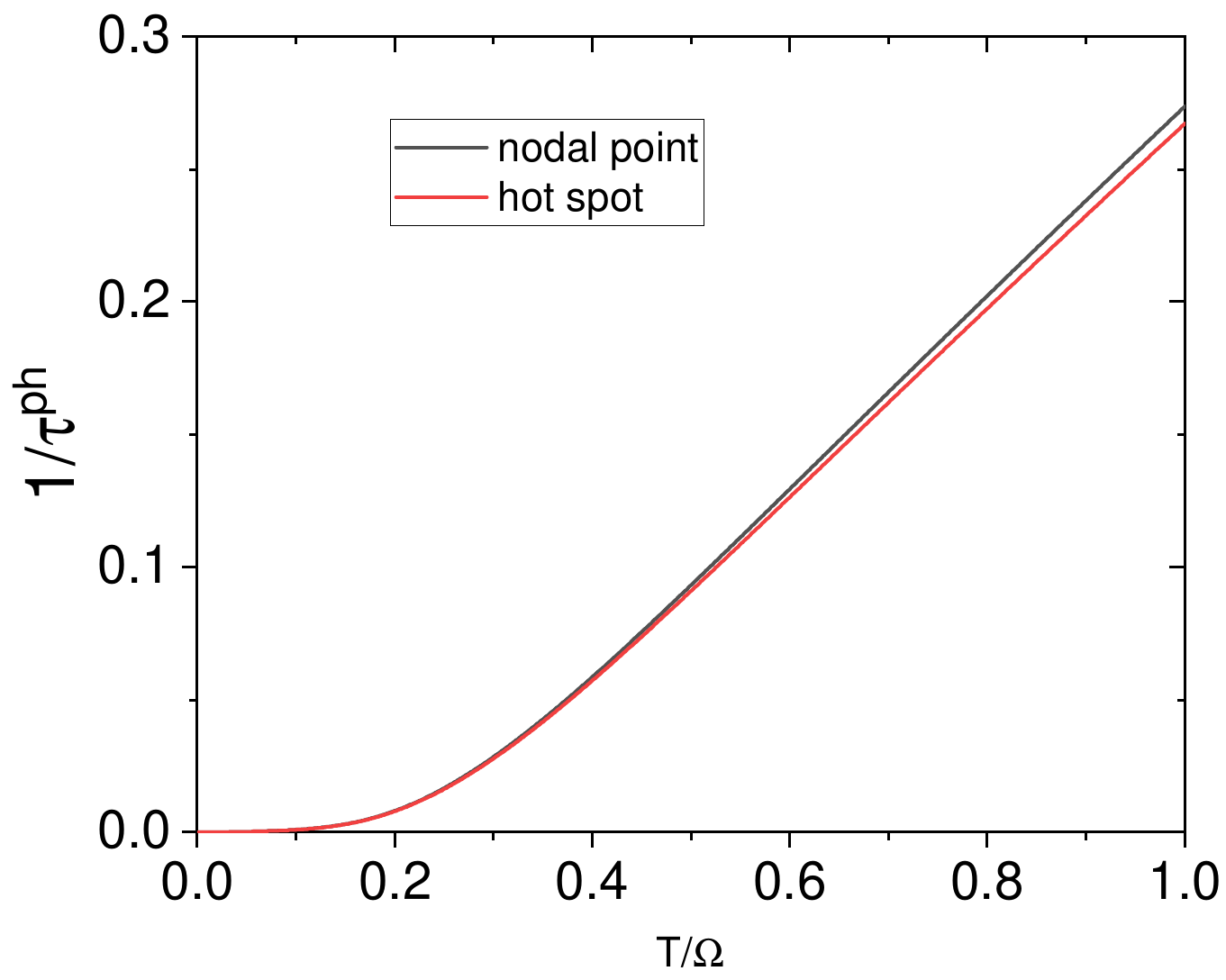}
\caption{The electron scattering rate at the fermi energy contributed by electron-phonon coupling. $1/\tau^{ph}$ is found to be almost isotropic on the fermi surface. Shown here is the scattering rate at the nodal point and the hot spot. The scattering rate is exponentially suppressed at low temperature and becomes approximately linear in T for $T\ge 0.5\Omega$.}
\end{figure}  

 \begin{figure}
\includegraphics[width=8cm]{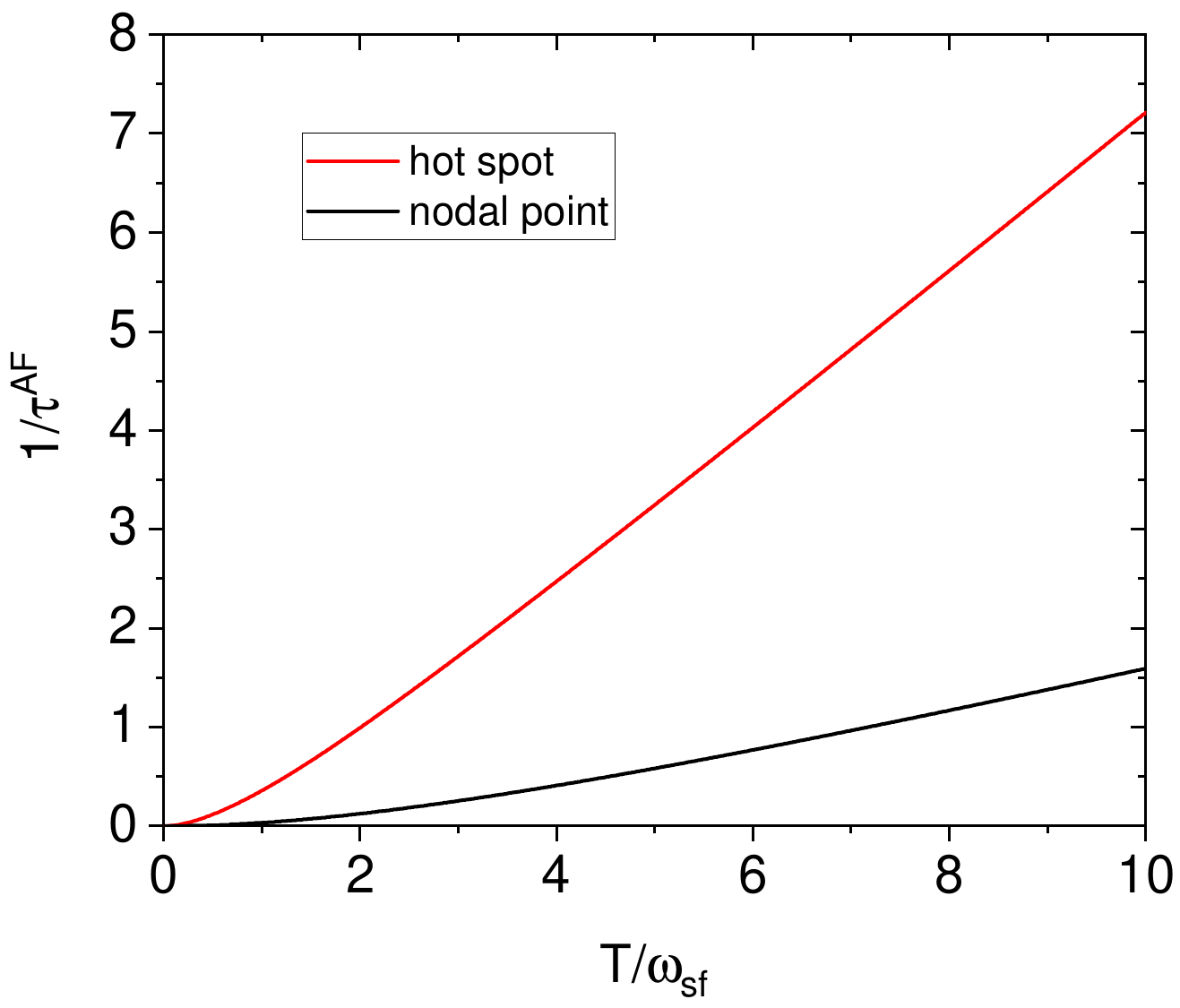}
\caption{The electron scattering rate at the fermi energy contributed by the antiferromagnetic spin fluctuation. $1/\tau^{AF}$ is seen to be strongly anisotropic on the fermi surface. As its name would suggest, the scattering rate at the hot spot is much larger than that at the nodal point(also called the cold spot in the literature). The scattering rate exhibits quadratic temperature dependence for $T\le \omega_{sf}(\mathbf{k})$, with $ \omega_{sf}(\mathbf{k})$ a momentum dependent characteristic temperature scale. $1/\tau^{AF}$ crossovers slowly to linear temperature dependence at higher temperature. Perfect linear-in-T behavior is only realized at a temperature significantly higher than $\omega_{sf}(\mathbf{k})$. The coupling strength is set to be $g=t$ in the calculation.}
\end{figure}  

\begin{figure}
\includegraphics[width=8cm]{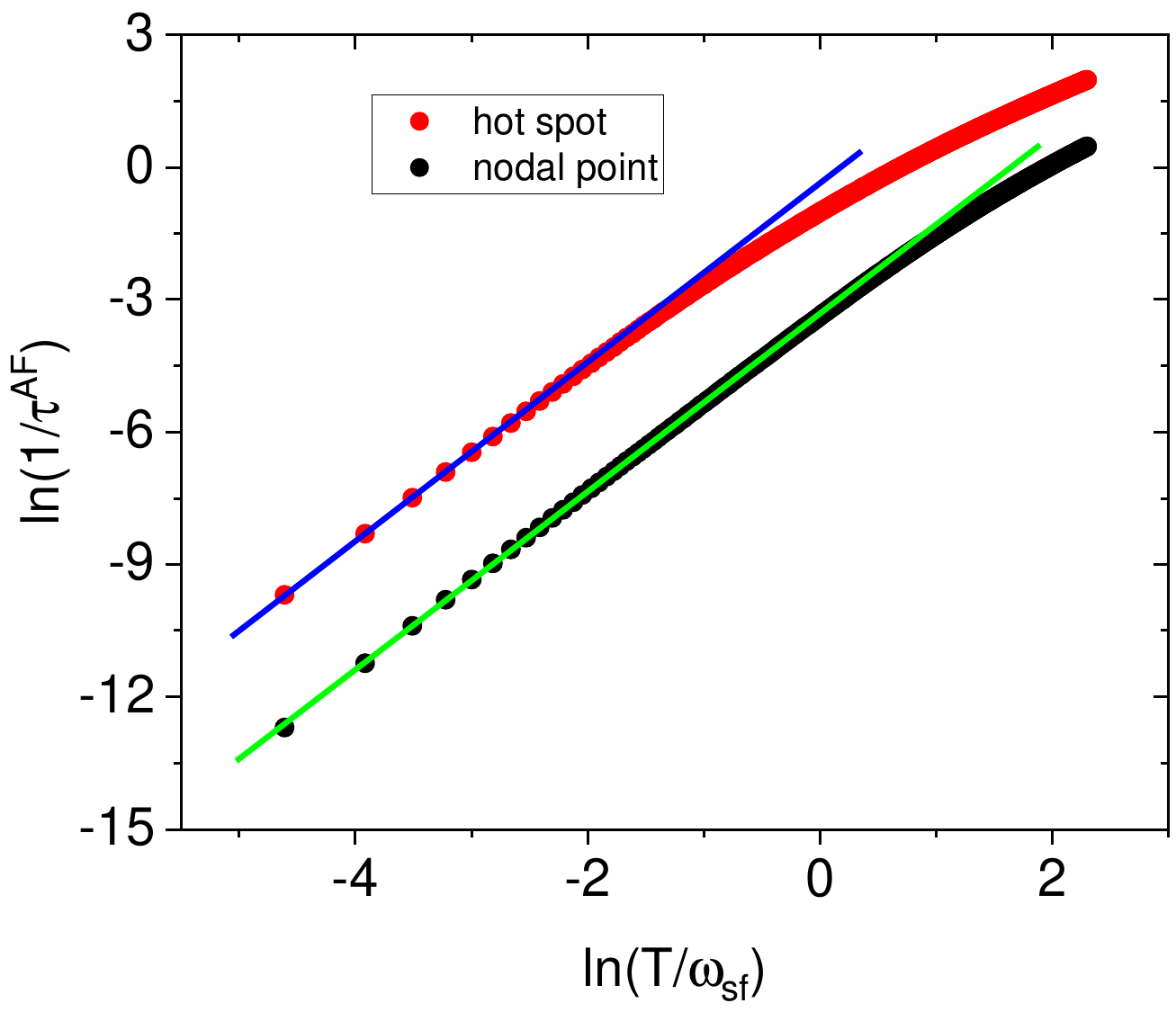}
\caption{The quadratic temperature dependence of the electron scattering rate $1/\tau^{AF}$ contributed by the antiferromagnetic spin fluctuation at low temperature. Perfect quadratic temperature dependence of $1/\tau^{AF}$ is seen below some momentum dependent characteristic frequency $\omega_{sf}(\mathbf{k})$ at both the nodal point and the hot spot. The blue and the green solid lines with slope 2 are guide to the eyes. However, the value of $\omega_{sf}(\mathbf{k})$ at the nodal point is clearly much larger than that at the hot spot. Correspondingly, the magnitude of $1/\tau^{AF}$ at the hot spot is much larger than that at the nodal point. The coupling strength is set to be $g=t$ in the calculation.}
\end{figure}  

We now calculate the quasiparticle scattering rate at the fermi energy as a function of temperature, which is given by
\begin{equation}
\frac{1}{\tau}=-\mathbf{Im} \Sigma^{r}(\mathbf{k}_{F},i0^{+})
\end{equation}
Here $\mathbf{k}_{F}$ denotes the fermi momentum. The scattering rate $1/\tau^{ph}$ caused by the electron-phonon coupling is shown in Fig.4. $1/\tau^{ph}$ is seen to be almost isotropic on the whole fermi surface. For $T\ll \Omega$, the scattering rate is exponentially suppressed as there is no thermally excited phonon. The scattering rate increases monotonically with temperature and gradually approaches a perfect linear temperature dependence for $T\ge 0.5\Omega$. 

The scattering rate $1/\tau^{AF}$ caused by the antiferromagnetic spin fluctuation is found to be strongly anisotropic on the fermi surface. Shown in Fig.5 is the results for the nodal point and the hot spot.  $1/\tau^{AF}$ is found to exhibit the standard quadratic temperature dependence below a momentum dependent characteristic frequency $\omega_{sf}(k)$(see Fig.6).  $\omega_{sf}(k)$ approaches its minimum along the fermi surface at the hot spots and its maximum at the nodal point. At higher temperature,  $1/\tau^{AF}$ gradually evolves into the linear temperature dependence. However, we note that a perfect linear-in-T behavior only appear at a temperature significantly higher than $\omega_{sf}(k)$. For example, we find that deviation from the linear-in-T behavior is still observable even at a temperature as high as $10\omega_{sf}$ at the nodal point. Since the reported $\omega_{sf}$ in the superconducting cuprates never reach below $4\ meV$\cite{Zhu}, it seems unlikely that the AF scattering scenario alone to be able to explain the linear-in-T resistivity below $300\ K$ in the cuprate superconductors in the perturbative scheme.

 \begin{figure}
\includegraphics[width=8cm]{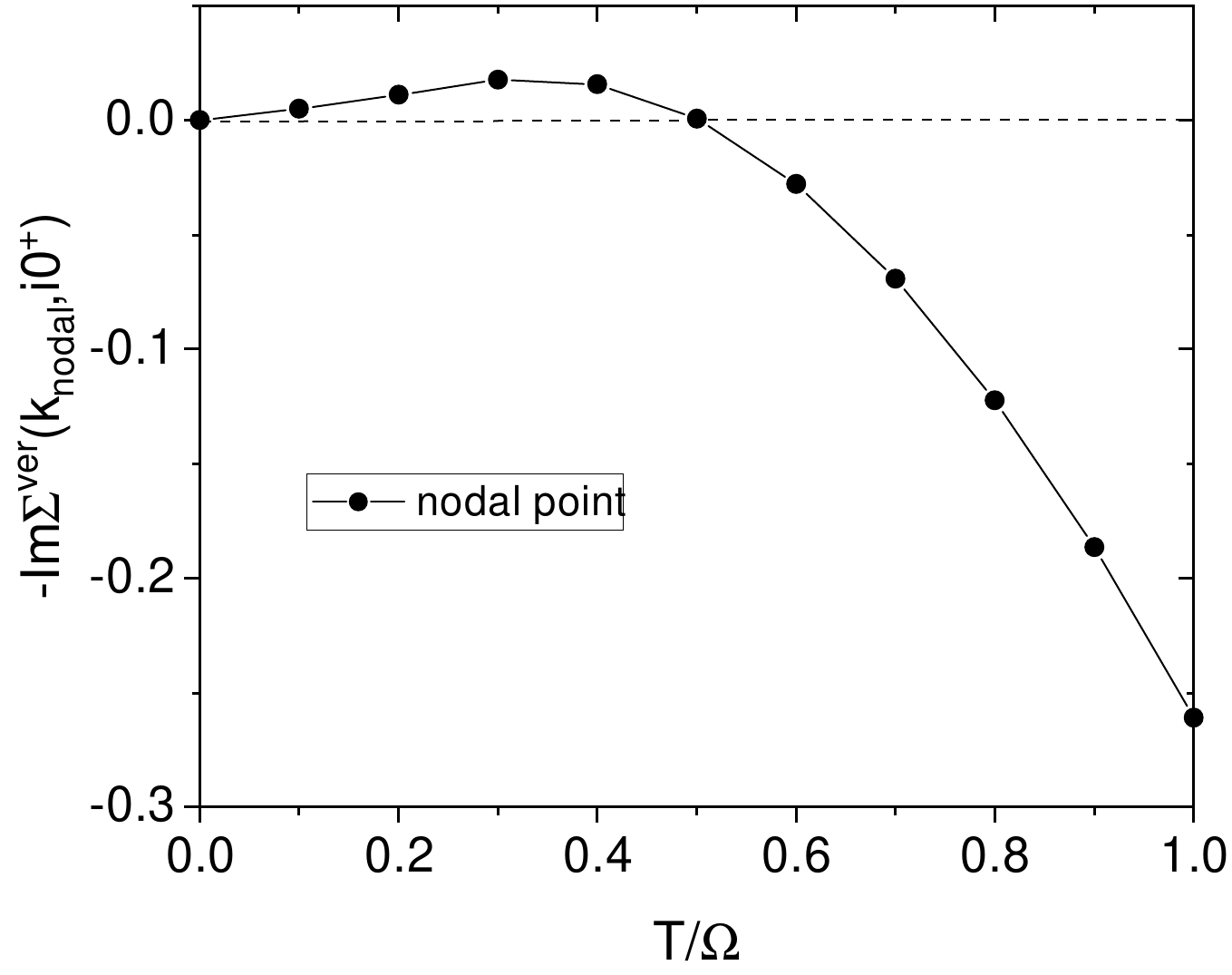}
\caption{The intertwined contribution to the electron scattering rate from both the $B_{1g}$ phonon and the antiferromagnetic spin fluctuation at the nodal point. A small and positive contribution with an approximate quadratic temperature dependence is observed at low temperature. $1/\tau^{ver}=-\mathrm{Im}\Sigma^{ver}$ becomes negative above $T=0.5\Omega$ and then increases rapidly in magnitude with $T$, reaching a value that is almost identical to that of $1/\tau^{ph}$ around $T=\Omega$. This implies that the destructive interference effect of the antiferromagnetic spin fluctuation on the electron-phonon coupling tends to cancel out the phonon contribution to the electron scattering rate. The coupling strength is set to be $g=t$ in the calculation.}
\end{figure}  

\begin{figure}
\includegraphics[width=8cm]{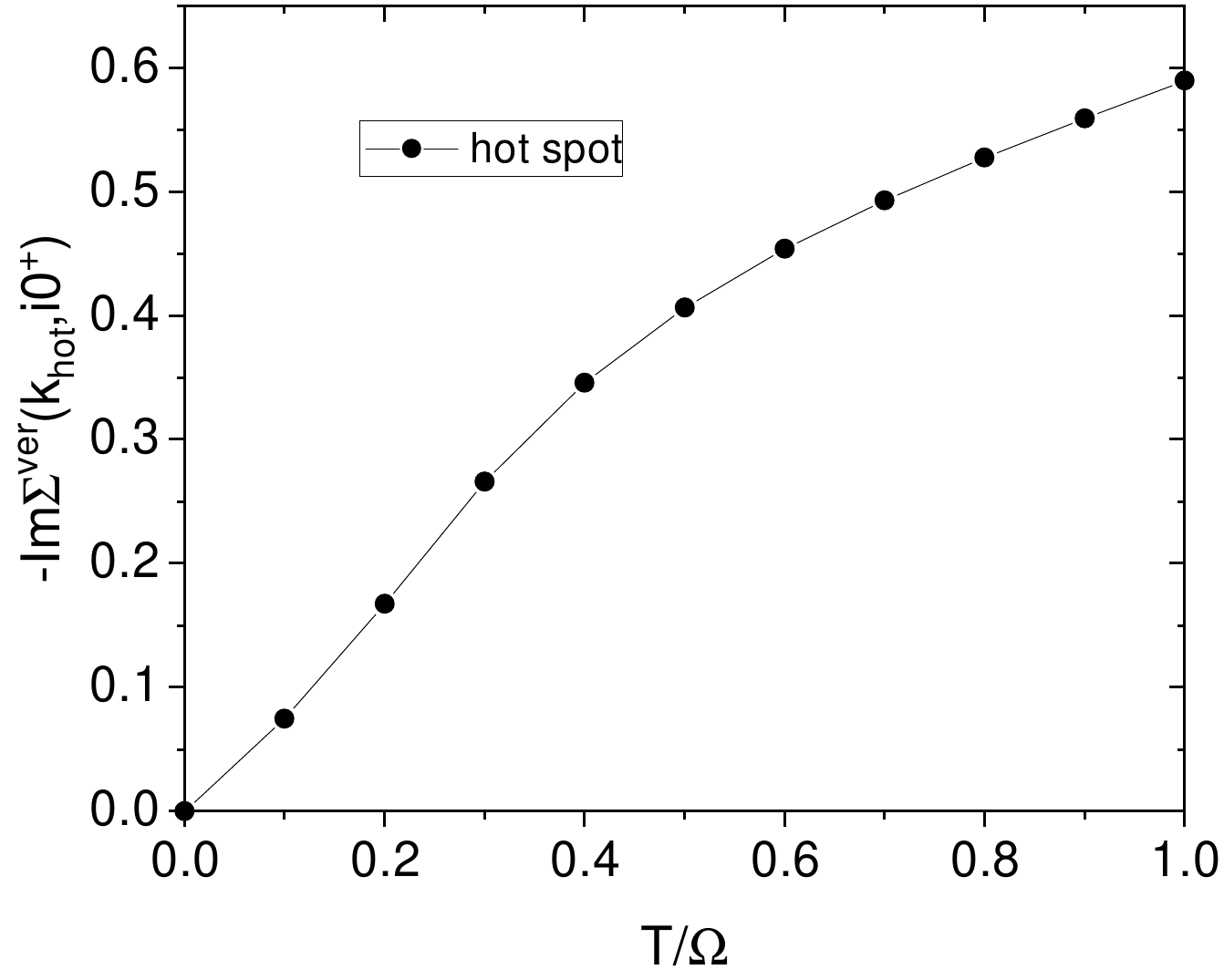}
\caption{The intertwined contribution to the electron scattering rate from both the $B_{1g}$ phonon and the antiferromagnetic spin fluctuation at the hot spot. A positive contribution with an approximate linear temperature dependence is observed below $T=0.5\Omega$. This is followed by another linear regime with a reduced slope through a kink structure around $T=0.5\Omega$. As is discussed in the main text, the origin of the kink structure can also be related to the destructive interference effect of the antiferromagnetic spin fluctuation on the electron-phonon vertex. The coupling strength is set to be $g=t$ in the calculation.}
\end{figure}

\begin{figure}
\includegraphics[width=8cm]{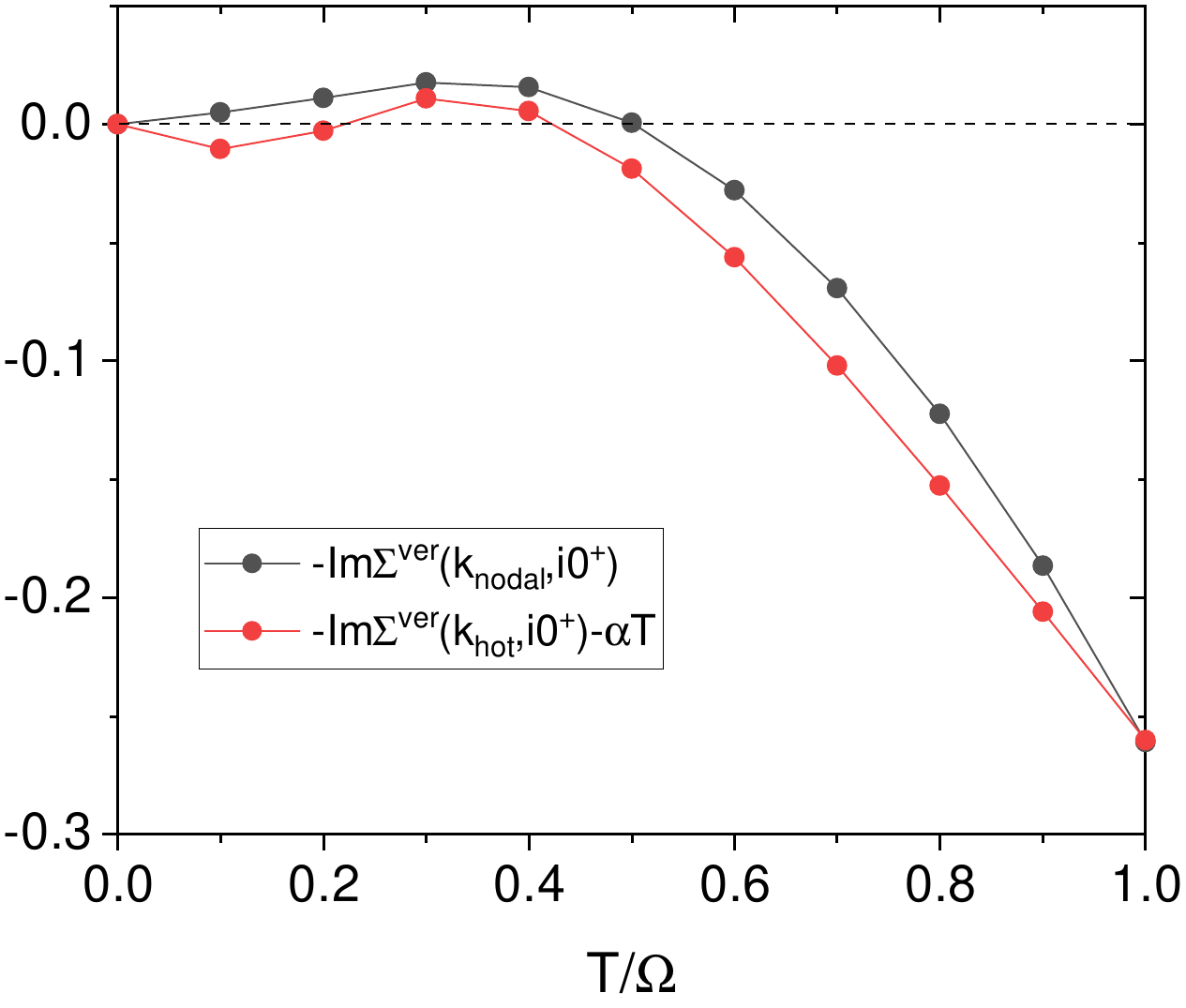}
\caption{Comparison between the temperature dependence of the intertwined contribution to the electron scattering rate at the nodal point and the hot spot. A linear temperature dependent background of the form $\alpha T$ is subtracted from $-\mathbf{Im}\Sigma^{ver}(\mathbf{k}_{hot},i0^{+})$. $\alpha$ is determined by a linear fit of $-\mathbf{Im}\Sigma^{ver}(\mathbf{k}_{hot},i0^{+})$ below $T=0.5\Omega$. Such a low temperature linear contribution is believed to be caused by the scattering of antiferromagnetic spin fluctuation. The coupling strength is set to be $g=t$ in the calculation.}
\end{figure}

We now turn to $1/\tau^{ver}=-\mathrm{Im}\Sigma^{ver}$, the contribution to the electron scattering rate from the intertwining term $\Sigma^{ver}$. We note that such a contribution can be  interpreted either as the vertex correction to $\Sigma^{ph}$ from the antiferromagnetic spin fluctuation, or, as the vertex correction to $\Sigma^{AF}$ from the $B_{1g}$ phonon. The electron scattering rate contributed by the intertwining term at the nodal point is shown in Fig.7.  It is small and positive for $T\le 0.5\Omega$ and exhibits an approximate quadratic temperature dependence. This is understandable since in this temperature regime the $B_{1g}$ phonon is not sufficiently excited and the main contribution to the electron scattering rate is from low energy spin fluctuation. $1/\tau^{ver}$ becomes significant and negative when $T\ge 0.5 \Omega$. In particular, its magnitude becomes almost identical to $1/\tau^{ph}$ around $T=\Omega$. A natural explanation of this result is that the destructive interference effect in the vertex correction from the antiferromagnetic spin fluctuation tends to cancel out the phonon contribution to the electron scattering rate. 

The intertwining contribution $1/\tau^{ver}$ becomes more subtle at the hot spot. As is shown in Fig.8, $1/\tau^{ver}$ is always positive at the hot spot, albeit with a kink structure around $T=0.5\Omega$, which separates an approximate linear regime at lower temperature from another linear regime with a reduced slope at higher temperature. We believe that the linear behavior below the kink should still be attributed to the scattering by thermally excited antiferromagnetic spin fluctuation. To understand the origin of the kink, we subtract from $1/\tau^{ver}$ a linear background of the form $\alpha T$. The result is then compared to $1/\tau^{ver}$ evaluated at the nodal point in Fig.9. It is remarkable to note that after the subtraction of the linear background $1/\tau^{ver}$ at the hot spot becomes essentially identical to that at the nodal point, apart from the small difference at low temperature(such a difference at low temperature is expected since $-\mathrm{Im}\Sigma^{ver}$ deviates from linear temperature dependence at low temperature at both the hot spot and the nodal point). It is thus plausible to interpret the kink structure around $T=0.5\Omega$ at the hot spot also as the consequence of the destructive interference effect in the vertex correction of the electron-phonon coupling by the antiferromagnetic spin fluctuation. We thus conclude that at the temperature scale of the phonon frequency, the intertwining term acts isotropically on the whole fermi surface to suppress the effective electron-phonon coupling. For the parameter we have chosen($g=t$), the phonon contribution to the electron scattering rate $1/\tau^{ph}$ is almost exactly canceled by the intertwining term.

From these discussions, we see that the coupling to the $B_{1g}$ oxygen buckling mode is strongly suppressed by the antiferromagnetic spin correlation in the cuprate superconductors. To fully address the puzzle concerning the violation of the Matthiessen's rule in the strange metal phase, we should also take into account the contribution to the electron scattering rate from other phonon modes. As we discussed in the introduction, the electron-phonon coupling in the cuprate superconductors can be roughly classified into two categories, namely, the diagonal coupling to the local electron density and the off-diagonal coupling to electron hopping terms. What we have shown here is that the off-diagonal coupling to nearest-neighboring hopping term for the $B_{1g}$ oxygen buckling mode is suppressed in an antiferromagnetic correlated background. Since the diagonal coupling to the local electron density is believed to be suppressed by the Mott physics\cite{Nagaosa}, the electron-phonon coupling effect in the cuprate superconductors is expected to be suppressed in the underdoped regime as a result of the proximity to the Mott insulating phase and the strong antiferromagnetic correlation in the system. This may provide a phenomenological understanding on why the Matthiessen's rule is violated in the strange metal phase of the cuprate superconductors. 

We note that a full account of the strange metal behavior is surely beyond the reach of the perturbative scheme adopted in this paper. According to our calculation, the electron scattering rate will eventually approach the fermi liquid regime at sufficient low temperature. Nevertheless, the intertwining between the electron-phonon coupling and the antiferromagnetic correlation between the electrons discussed in this work must be an integrated part of the full story of the strange metal physics. A non-perturbative account of the electron-phonon coupling and the strong correlation effect on the equal footing is a challenging task. However, we note that a recent development in the effective field theory of the Hubbard model may shade new light on the solution of such a hard problem\cite{Liu}.

Beside the electron scattering rate discussed in this work, the full frequency dependence of the electron self energy we obtained here can be exploited to analyze the spectral anomalies observed in the cuprate superconductors, for example, the dispersion kink in the nodal direction and the peak-dip-hump structure in the anti-nodal region. Historically, it had been strongly debated if we should attribute these spectral anomalies to the electron-phonon coupling or the coupling to some mode of electronic origin, in particular, the neutron resonance mode\cite{He,Damascelli,Louie,Reznik,Heid,Dessau,Li,Zhou}. The argument presented in this work suggests that the electron-phonon coupling may not be the principle origin of these spectral anomalies in the underdoped regime. A detailed analysis on this issue will be pursued in a separate work.

In summary, we have shown that the strong vertex correction from the antiferromagnetic spin fluctuation tends to suppress the coupling between the electron and the $B_{1g}$ buckling phonon mode in the cuprate superconductors. According to our scenario, the effective electron-phonon coupling strength should $decrease$ with the proximity of the Mott insulating phase and the resultant enhancement of the antiferromagnetic correlation in the underdoped cuprate superconductors.  We argue that the spectral anomalies observed in the underdoped cuprates, especially the peak-dip-hump structure observed in the anti-nodal region, should not be attributed to the electron-phonon coupling. We propose that the absence of phonon signature in the dc resistivity of the strange metal phase, namely, the violation of the Matthiessen's rule, should be attributed to the destructive interference effect in the vertex correction to the electron-phonon coupling by the antiferromagnetic spin fluctuation in the system.

 \begin{acknowledgments}
We acknowledge the support from the grant NSFC 12274453.
\end{acknowledgments}

\end{document}